\documentclass{article}

\usepackage[utf8]{inputenc}
\usepackage{tismir}
\usepackage{amsmath}
\usepackage{hyperref}
\usepackage{url}
\usepackage{amssymb}
\usepackage{graphicx}
\usepackage{caption}
\usepackage{booktabs}
\usepackage{multirow}
\usepackage{xcolor}
\usepackage{tabularx}
\usepackage{booktabs}
\usepackage[table]{xcolor}
\usepackage{CJKutf8}
\usepackage{float}
\title{HAIM: Human-AI Music Datasets for AI Music Production Tracking Benchmark}
\author{Seonghyeon Go, Yumin Kim}
\date{}

\authorref{Go \& Kim}
\authorshort{Go \& Kim}

\begin{document}

\twocolumn[{%
\maketitleblock
\begin{abstract}
As generative platforms such as Suno and Udio reach human-grade audio quality, the scope of AI's utility has expanded across the entire music production workflow. Beyond simple track generation, these advancements have catalyzed the adoption of AI-driven methodologies in diverse forms. These include vocal synthesis, arrangement, and professional mastering. However, current detection research remains largely confined to a binary `AI-or-human' paradigm. It fails to reflect the realities of contemporary music production workflows. In real-world production, AI tools are increasingly used to refine or master human-produced tracks, and human engineers likewise post-process AI-generated material to ensure professional quality. Moreover, users often employ adversarial tactics to bypass AI detectors, such as applying human mastering to AI-generated tracks. This creates a grey area that a simple binary classification fails to capture. In this paper, we define and investigate ``AI Music Tracking'': the challenge of identifying specific AI integration across the multifaceted spectrum of music production. To this end, we introduce HAIM, a dataset with diverse labels for stages of music production. It is designed to isolate stages of AI intervention, including hybrid production and agent-level tracking. Our evaluation of state-of-the-art detectors reveals systemic flaws. By releasing HAIM, we propose a new benchmark that shifts the field beyond binary classification toward a granular, structured evaluation of AI music.

\end{abstract}
}]
\saythanks{}

%===============================================================================
\section{Introduction}\label{sec:intro}
%===============================================================================

Since late 2023, the generative music landscape has undergone a paradigm shift. Commercial platforms such as Suno~\citep{suno2024} and Udio~\citep{udio2024} now produce studio-quality tracks with sophisticated vocal rendering, while open-source architectures like ACE-Step-1.5~\citep{gong2026acestep} have demonstrated that high-quality synthesis is achievable on consumer-grade hardware. The resulting influx of AI-generated content is substantial: according to a recent industry report~\citep{deezer2025sep}, fully AI-generated tracks account for over 28\% of daily uploads to Deezer.

The detection community has responded with progressively larger and more diverse benchmarks.
SONICS~\citep{rahman2025sonics} introduced 97k full-song pairs alongside the SpecTTTra architecture, reported to be 38\% faster and 26\% more memory-efficient than standard Vision Transformers.
AIME~\citep{grotschla2025benchmarking} broadened source diversity to 12 generation models, while MoM~\citep{crosvila2025momclam} scaled to 130k tracks with rigorous out-of-distribution (OOD) splits and the CLAM dual-stream detector.
On the detection task, systems from IRCAM Amplify~\citep{ircamamplify2024} and Deezer report near-perfect accuracy (up to 99.8\%) on known generators.
While binary detection, classifying a track as either entirely human or entirely synthetic, appears close to being solved, it addresses an increasingly incomplete picture of real-world production.

In modern workflows, AI is no longer treated as a monolithic source of audio; instead, it is integrated as a granular component across the production pipeline.~\citep{kim2025generation} From initial composition to automated mixing, the music value chain has evolved into a multi-stage hybrid process in which producers may incorporate AI-generated stems. Distinct from these artistic directions, some AI music producers reportedly apply manual mastering to avoid music platform detection systems. By processing synthetic audio with standardized plugin presets, these users seek to obscure its origin. These diverse practices, spanning production and evasion, blur the boundary between human and AI-generated content.

\begin{figure}[t]
\centering
\includegraphics[width=\linewidth]{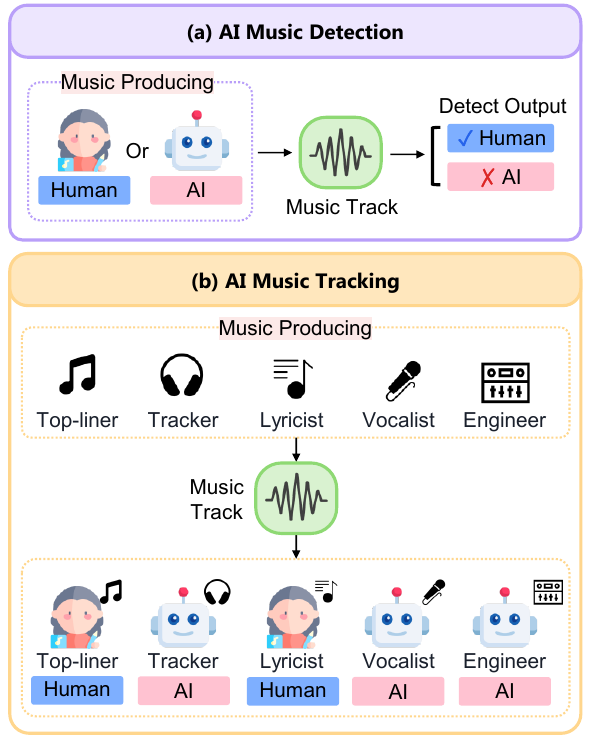}
\caption{Comparison of \textbf{AI Music Detection} and \textbf{AI Music Tracking}. Detection focuses on whole-track binary classification, whereas tracking operationalizes detail analysis to identify the nuances of AI integration. Figure shows the one case of analysis, multilabel classification.}
\label{fig:detection_vs_tracking}
\end{figure}

We argue that binary detection—classifying a track as either entirely human or entirely synthetic—is fundamentally misaligned with the reality of modern music production. In contemporary workflows, producers may combine human and AI elements. This hybridization can manifest in various ways: (1) AI vocals over human instrumentals, (2) AI-driven mastering of human demos, or (3) AI segments spliced between human passages to obscure transitions. The result is a fluid spectrum of authorship that no binary label can faithfully represent, necessitating a more granular tracking approach.

To make this mismatch precise, we distinguish two tasks (Figure~\ref{fig:detection_vs_tracking}). \textbf{AI Music Detection} treats provenance as a binary attribute, either human or machine, assuming each track has a single, unambiguous origin. When this assumption holds, existing systems perform remarkably well, with commercial solutions reporting accuracy exceeding 99\%~\citep{ircamamplify2024}. \textbf{AI Music Tracking} relaxes this assumption: rather than asking whether AI was totally involved, it asks how, where, and to what extent, decomposing contributions across components, production stages, and temporal segments. By shifting the focus from simple categorization to detailed provenance, HAIM moves beyond existing research trends in music classification and generation, pioneering ``AI Music Tracking'' as a distinct and critical problem domain for the generative era.

The need for AI tracking is not merely academic. Legal cases, including a musician charged with defrauding streaming platforms of over $10$ million via AI-generated bots, and emerging regulations such as the EU AI Act demand fine-grained provenance. However, current detection frameworks offer no mechanism for quantifying the \emph{degree} of AI involvement, precisely because they were never designed to. Closing this gap requires both a new task formulation and benchmarks that capture the full spectrum of human-AI collaboration.

To address these limitations, we introduce \textbf{HAIM} (\textbf{H}uman--\textbf{AI} \textbf{M}usic Dataset), a benchmark designed to evaluate tracking capabilities across the full creative spectrum. In this work, we use the term \emph{AI music} to refer to full-length tracks or stems generated offline via text prompting or reference conditioning, as opposed to real-time or interactive co-creative systems. In addition, we train and observe a tracking model with various perspectives based on the existing AI detection model. Our contributions are as follows:
\begin{enumerate}[leftmargin=*,itemsep=2pt,topsep=2pt]
    \item \textbf{Multi-faceted Taxonomy.}\; We organize 196,000 tracks into 13 categories across three groups, isolating specific stages of AI integration---from pure baselines through component-level hybridization to temporal splicing.
    \item \textbf{Diverse Data Sourcing.}\; We curate tracks from various generation backbone models spanning multiple architectural paradigms, and by various methods like manual generation, crowd sourcing, or post-processing.
    \item \textbf{Comprehensive Benchmarking.}\; We evaluate four open-source detection systems on every category, revealing systematic failure modes in hybrid scenarios.
    \item \textbf{Tracking-capable Detector.}\; We trained MuQ-based~\citep{zhu2025muq} Fusion Segment Transformer~\citep{kim2026fusion} that performs both binary detection and multi-label AI music tracking. By predicting AI involvement across four key production roles—Composer, Lyricist, Vocalist, and Audio Engineer—our model provides a structured characterization that operationalizes the AI tracking task.
\end{enumerate}

% =============================================================================
% TABLE 1: Dataset Comparison (revised — role-level decomposition as key axis)
% =============================================================================

\begin{table*}[t]
\centering

\small
\renewcommand{\arraystretch}{1.15}
\begin{tabularx}{\textwidth}{X|ccccc|c}
\toprule
\textbf{Datasets} & \textbf{SONICS} & \textbf{AIME} & \textbf{MoM} & \textbf{OpenBMAT} &\cite{cros2025ai} & \textbf{HAIM (Ours)} \\
\midrule
\textbf{Music-level decomposition}          &   &      &   & & &  \\
\quad Temporal mix & --        & --        & --        & --        & --        &  \checkmark\\
\quad Multi-generator sources           & $\triangle$   & \checkmark & \checkmark & --        & \checkmark   & \checkmark \\
\midrule
\textbf{Role-level decomposition}          &   &      &   & & &  \\
\quad Composer       & --        & --        & --        & --        & --        & \checkmark \\
\quad Lyricist                    & $\triangle$   & --        & $\triangle$   & --        & --        & \checkmark \\
\quad Vocalist                    & --        & --        & $\triangle$   & --        & --        & \checkmark \\
\quad Audio Engineer (mix\&master)  & --        & --        & --        & --        & --        & \checkmark \\
\midrule
Total tracks       & 97K  & 6.5K & 130K & 3.3K & 30K & 196K \\
\bottomrule
\end{tabularx}
\caption{\textbf{Comparison of HAIM with existing AI-generated music detection datasets.}
\checkmark\,=\,supported, $\triangle$\,=\,partial, --\,=\,absent.
HAIM is the first benchmark to decompose AI involvement along music production roles.}
\label{tab:dataset_comparison}
\end{table*}

%===============================================================================
\section{Related Work}\label{sec:related}
%===============================================================================

\paragraph{Detection systems.}
On the commercial front, IRCAM Amplify~\citep{ircamamplify2024} reports accuracy up to 99\%, while Deezer's system claims 99.8\% by identifying neural decoder artifacts, though the authors themselves demonstrate that this figure is fragile under pitch shifting, noise injection, and codec re-encoding~\citep{afchar2024detecting}. CLAM~\citep{crosvila2025momclam} represents a notable advance, leveraging parallel MERT~\citep{li2024mert} and Wav2Vec~\citep{baevski2020wav2vec} encoders fused via a Weighted Cross-Aggregation module to exploit statistical inconsistencies between vocal and instrumental streams. Deezer reports that spectral analysis on AI-generated content could give proper decisions~\citep{afchar2025detecting, afchar2025fourier}.
% While CLAM achieves an F1-score of 0.993 on SONICS, its sensitivity to the partial and layered AI involvement characteristic of modern production remains uninvestigated.
Fusion Segment Transformer (FST)~\citep{kim2026fusion} introduces a dual-stream architecture for full-audio detection: an embedding stream processes segment-level content features extracted by pretrained models, while a self-similarity matrix (SSM) stream captures repetitive structural patterns, with the two streams integrated via bi-directional cross-attention and a gated fusion mechanism. FST achieves state-of-the-art results on SONICS~\citep{rahman2025sonics} and AIME~\citep{grotschla2025benchmarking}. However, like all prior detectors, it outputs a single binary label and does not address multilabel role-level tracking. 

\paragraph{Detection datasets.}
Early research on AI-generated music detection focused primarily on Singing Voice Deepfake Detection (SVDD), employing benchmarks such as SingFake~\citep{zang2024singfake} to identify synthetic vocals superimposed on authentic instrumentals.
As generative models matured, AIME~\citep{grotschla2025benchmarking} introduced dataset of 6k tracks with 12 generation sources. And SONICS~\citep{rahman2025sonics} introduced a large-scale full-song dataset of 97k tracks with binary labeling and the SpecTTTra detector. However, its reliance on generative platforms (predominantly Suno~\citep{suno2024}, with some Udio~\citep{udio2024}) limits generalizability to the broader landscape of modern generators~\citep{cros2025ai}.
Subsequent efforts broadened the landscape: MoM~\citep{crosvila2025momclam} introduced 130{,}435 tracks with human covers as ``hard negatives'' and rigorous out-of-distribution (OOD) evaluation splits, in which generators used at test time are entirely absent from training, forcing detectors to generalize beyond memorized artifacts.
Despite these advances, all existing benchmarks assume an all-or-nothing involvement of AI, leaving a critical void in forensic tools for hybrid production workflows.

\paragraph{Generation systems.}
The generative landscape spans commercial leaders, Suno, Udio, Lyria-Pro, and Mureka~\citep{mureka2024} and open-source alternatives such as ACE-Step-1.5~\citep{gong2026acestep}.
Suno's evolution from v2 to v5 demonstrates a rapid reduction in detectable artifacts, progressing from audible chorus-like distortions to studio-quality audio.
ACE-Step-1.5 is particularly notable for its decoupled architecture, pairing a language model planner with a Diffusion Transformer (DiT)~\citep{peebles2023scalable}, which supports editing capabilities including repainting (inpainting) and variation.

% By including models spanning autoregressive, diffusion, and hybrid architectures, HAIM ensures its benchmark is not biased toward any single synthesis paradigm.

% =============================================================================
% TABLE 2: Production-Role Taxonomy (updated to match current folder structure)
% =============================================================================

\begin{table*}[t]
\centering
\small
\renewcommand{\arraystretch}{1.2}
\begin{tabular}{l l | c c c c | r}
\toprule
\textbf{Group} & \textbf{Category} & \textbf{Composer} & \textbf{Lyricist} & \textbf{Vocalist} & \textbf{Engineer} & \textbf{Tracks} \\
\midrule
\multirow{2}{*}{A: Baselines set}
  & A1 --- Human-composed real music              & H    & H    & H    & H    & 94{,}654 \\
  & A2 --- Full AI Generation      & AI   & AI   & AI   & AI   & 55{,}485 \\
\midrule
\multirow{4}{*}{\shortstack[l]{B: Mastering\\Hybrids}}
  & B1 --- AI-Mastered Human Track & H    & H    & H    & AI   & 6{,}000 \\
  & B2 --- AI Music \& Human-Ref.\ Mix & AI   & AI   & AI   & AI\textsuperscript{$\dagger$}    & 6{,}000 \\
  & B3 --- AI Music \& Human Mastering   & AI   & AI   & AI   & H    & 6{,}000 \\
  & B4 --- AI Music \& Human Mix         & AI   & AI   & AI   & H    & 6{,}000 \\
\midrule
\multirow{2}{*}{\shortstack[l]{B: Component\\Hybrids}}
  & B5 --- AI Vocal Cover          & H    & H    & AI   & AI   & 2{,}040 \\
  & B6 --- Human Lyrics w/ AI Gen   & AI   & H    & AI   & AI   & 2{,}000 \\
\midrule
\multirow{3}{*}{\shortstack[l]{B: AI-driven\\ Reference Editing}}
  & B7 --- AI Variation            & AI\textsuperscript{$\dagger$} & AI\textsuperscript{$\dagger$} & AI\textsuperscript{$\dagger$} & AI\textsuperscript{$\dagger$} & 2{,}000 \\
  & B8 --- AI Edit                 & AI\textsuperscript{$\dagger$} & AI\textsuperscript{$\dagger$} & AI\textsuperscript{$\dagger$} & AI\textsuperscript{$\dagger$} & 2{,}000 \\
  & B9 --- AI Repaint              & AI\textsuperscript{$\dagger$} & AI\textsuperscript{$\dagger$} & AI\textsuperscript{$\dagger$} & AI\textsuperscript{$\dagger$} & 2{,}000 \\
\midrule
\multirow{2}{*}{\shortstack[l]{C: Temporal\\Mix Set}}
  & C1 --- Concatenation           & H+AI & H+AI & H+AI & H+AI & 6{,}000 \\
  & C2 --- Crossfade               & H+AI & H+AI & H+AI & H+AI & 6{,}000 \\
\bottomrule
\end{tabular}
\caption{\textbf{HAIM category taxonomy mapped to music production roles.}
\textbf{H}\,=\,Human,
\textbf{AI}\,=\,AI-generated,
\textbf{AI\textsuperscript{$\dagger$}}\,=\,AI conditioned on or derived from human source material.
Unlike existing benchmarks that treat AI involvement as a binary whole-track property, the HAIM decomposes it along four production roles.}
\label{tab:role_taxonomy}
\end{table*}

%===============================================================================
\section{Dataset}\label{sec:dataset}
%===============================================================================
To provide a comprehensive framework for AI music detection, we introduce the HAIM (Human-AI Music) dataset for various benchmarks. As shown in Table~\ref{tab:dataset_comparison}, unlike existing datasets, it enables training and evaluation from multiple perspectives.

% , which is structured into two main parts. 
% First, Section~\ref{sec:design} describes the conceptual organization of thirteen categories across three groups, focusing on the collaborative roles between humans and AI agents. 
% Second, Section~\ref{sec:collection} details the practical implementation of these categories, including the specific AI models, web-scraping protocols, and professional engineering workflows used to generate the audio.

\subsection{Dataset Design}\label{sec:design}

Modern music production is a collaborative pipeline of specialized roles: a \emph{composer}---including a \emph{tracker} who architects the harmonic and rhythmic foundation, and a \emph{top-liner} who crafts the topline melody---constructs the primary musical framework, a \emph{lyricist} writes the lyrics, a \emph{vocalist} performs them, and an \emph{audio engineer} refines the output through mixing and mastering. While AI can now substitute any combination of these roles, existing benchmarks treat AI involvement as a binary property, overlooking the compositional nature of real-world AI-assisted production.

HAIM addresses this gap by organizing thirteen categories into three groups (Table~\ref{tab:role_taxonomy}). The two baseline groups (A1, A2) are drawn from large-scale sources and are therefore the largest, while the hybrid (Group B) and temporal-mix (Group C) categories rely on manual processing or single-generator pipelines and are consequently smaller; we regard scaling these categories as future work (Section~\ref{sec:discussion}).

\textbf{Group A (Baselines)} establishes the two extremes: fully human (A1) and fully AI (A2), with all production roles uniformly assigned to a single agent type.

\textbf{Group B (Hybrid Mix Set)} isolates specific production roles for AI substitution. Mastering-level hybrids (B1--B4) test whether post-production can obscure generation artifacts. Component-level hybrids (B5, B6) partition roles along semantic boundaries---AI vocals over human instrumentals, or human lyrics with AI generation. AI-driven reference editing (B7--B9) applies variation, edit, and repaint operations to human source material.

\textbf{Group C (Temporal Mix Set)} interleaves human and AI segments along the time axis via concatenation (C1) and crossfading (C2), evaluating robustness to mid-track distribution shifts.

\begin{figure*}[!t]
  \centering
  \includegraphics[width=\textwidth]{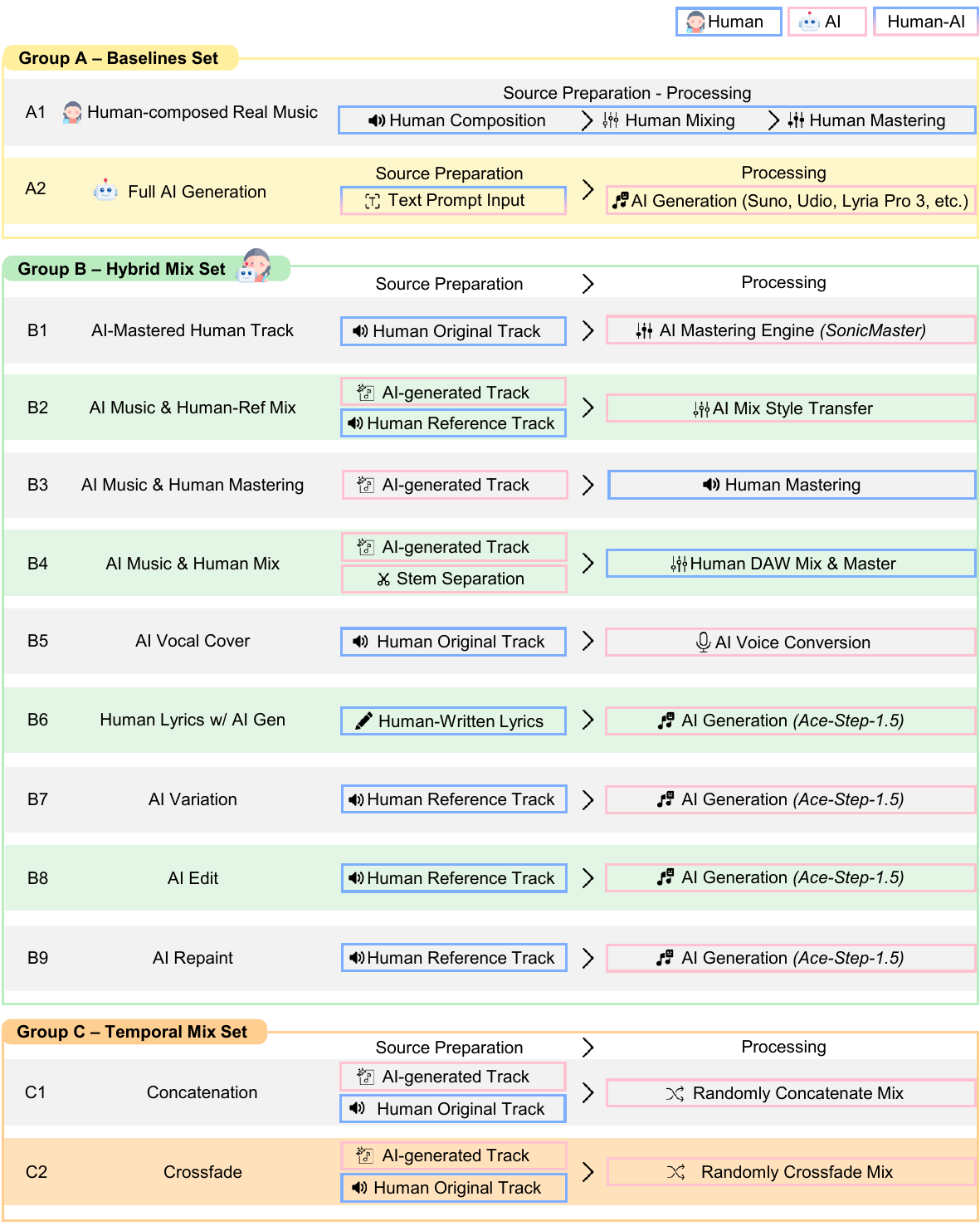}
  \caption{\textbf{Overview of the HAIM dataset generation pipeline.} The schematic details the sourcing and processing stages for each category: (Group A) establishes baselines from human and AI sources; (Group B) constructs hybrid tracks through role-based processing such as AI mastering (B1), style transfer (B2), and voice conversion (B5); and (Group C) generates temporal mixtures through segment-level concatenation (C1) and crossfading (C2).}
  \label{fig:pipeline}
\end{figure*}

%===============================================================================
\subsection{Data Sourcing and Generation}\label{sec:collection}
%===============================================================================
To reflect common real-world music production, we followed a typical studio production sequence (Figure~\ref{fig:pipeline}). The only exceptions are subsets of closed-source data in group \textbf{A} and category \textbf{B5}, which were constructed via web scraping from online platforms.

%--- A: Baselines ---%
\paragraph{A1: Human-composed Real Music.}
Human-produced tracks are sourced from the MTG-Jamendo dataset~\citep{bogdanov2019mtg}, providing 55,000 full-length tracks, and the SONICS dataset~\citep{rahman2025sonics}, contributing 44,654 YouTube tracks to ensure broad stylistic coverage.

\paragraph{A2: Full AI Generation.}
A2 incorporates 55,485 tracks from six generation tools. Four commercial platforms were collected via web scraping:

\begin{itemize}[nosep,leftmargin=*]
\item \textbf{Suno:} Following the ID-based collection strategy of \citep{cros2025ai}, we harvested song UUIDs from public feed APIs on CDN endpoints. Version-specific metadata—including v4, v4.5, v4.5+, and v5—was extracted directly from page HTML via \texttt{model\_version} and \texttt{model\_name} fields to ensure granular version control.

\item \textbf{Udio:} Data was collected via the platform's search API using genre-specific queries across three distinct sort modes. To ensure version diversity, the tracks were categorized into \texttt{v1}, \texttt{v1.5}, and \texttt{v2} based on model metadata.

\item \textbf{Mureka:} Data was sourced via sitemap parsing and API pagination, with versions \texttt{v6}, \texttt{v7.5}, and \texttt{v8} identified through model metadata fields.
    
\item \textbf{Lyria Pro 3:} Generated via Google DeepMind's API, this dataset provides a high-fidelity benchmark to challenge modern detection systems.
    
\item \textbf{MusicGen~\citep{copet2023musicgen}:} Locally generated on an NVIDIA RTX 5090 across three model scales (small, medium, large). All outputs are short-form music (10\,s or 30\,s) at 32\,kHz.
    
\item \textbf{ACE-Step-1.5:} Locally generated using 40 genre-diverse prompts authored with Gemini Pro 3. Its reproducible pipeline and modular modes—including \textit{edit}, \textit{variation}, and \textit{repaint}—serve as the technical foundation for Group B.
\end{itemize}

%--- B: Mastering Hybrids ---%

\paragraph{B1: AI-Mastered with Human Track.}
AI mastering is applied to human-produced tracks from A1 (Subset of MTG-Jamendo) using SonicMaster~\citep{melechovsky2025sonicmaster}, a flow-matching generative model conditioned on natural-language instructions. 

\paragraph{B2: AI Music with Human-reference Mix.}
A pre-trained FXencoder~\citep{koo2023music} extracts mixing-style embeddings---capturing EQ curves, dynamic range, and stereo imaging---from randomly selected human-produced reference tracks (MTG-Jamendo in A1). These embeddings are applied to ACE-Step-1.5-generated tracks, reshaping their spectral and dynamic profiles to match the human reference.

\paragraph{B3: AI Music with Human Mastering.}
A professional audio engineer performed stereo mastering on ACE-Step-1.5 tracks within FL Studio, using a two-stage chain: FabFilter Pro-Q 3 (steep bandpass, 20\, Hz--20\,kHz) and iZotope Ozone 9 (multiband dynamics, stereo imaging, harmonic excitation, brickwall limiting). Factory presets were used throughout to establish a reproducible upper bound on professional mastering's ability to obscure AI artifacts.

\paragraph{B4: AI Music with Human Mix.}
ACE-Step-1.5 tracks were decomposed into stems (vocals, bass, drums, other) via htdemucs~\citep{defossez2021demucs} source separation. A professional engineer performed per-stem mixing in FL Studio:
\begin{itemize}[nosep,leftmargin=*]
  \item \textbf{Vocals}: FabFilter Pro-R 2 (reverb) $\to$ Waves CLA Vocals $\to$ Waves DeEsser $\to$ FabFilter Pro-C 2.
  \item \textbf{Bass}: FabFilter Pro-C 2 $\to$ FabFilter Pro-Q 3 $\to$ Waves CLA Bass.
  \item \textbf{Drums}: FabFilter Pro-Q 3 $\to$ Waves Smack Attack $\to$ FabFilter Pro-C 2 $\to$ Waves DeEsser.
  \item \textbf{Other}: FabFilter Pro-C 2 $\to$ iZotope Ozone 9.
\end{itemize}
Stereo placement followed a fixed configuration applied uniformly across tracks: vocals widened by 10\%, bass narrowed by 15\% toward mono, and the `other' stem widened by 20\%. The stereo bounce was mastered identically to B3.

\begin{figure*}[!t]
  \centering
  \includegraphics[width=0.98\textwidth]{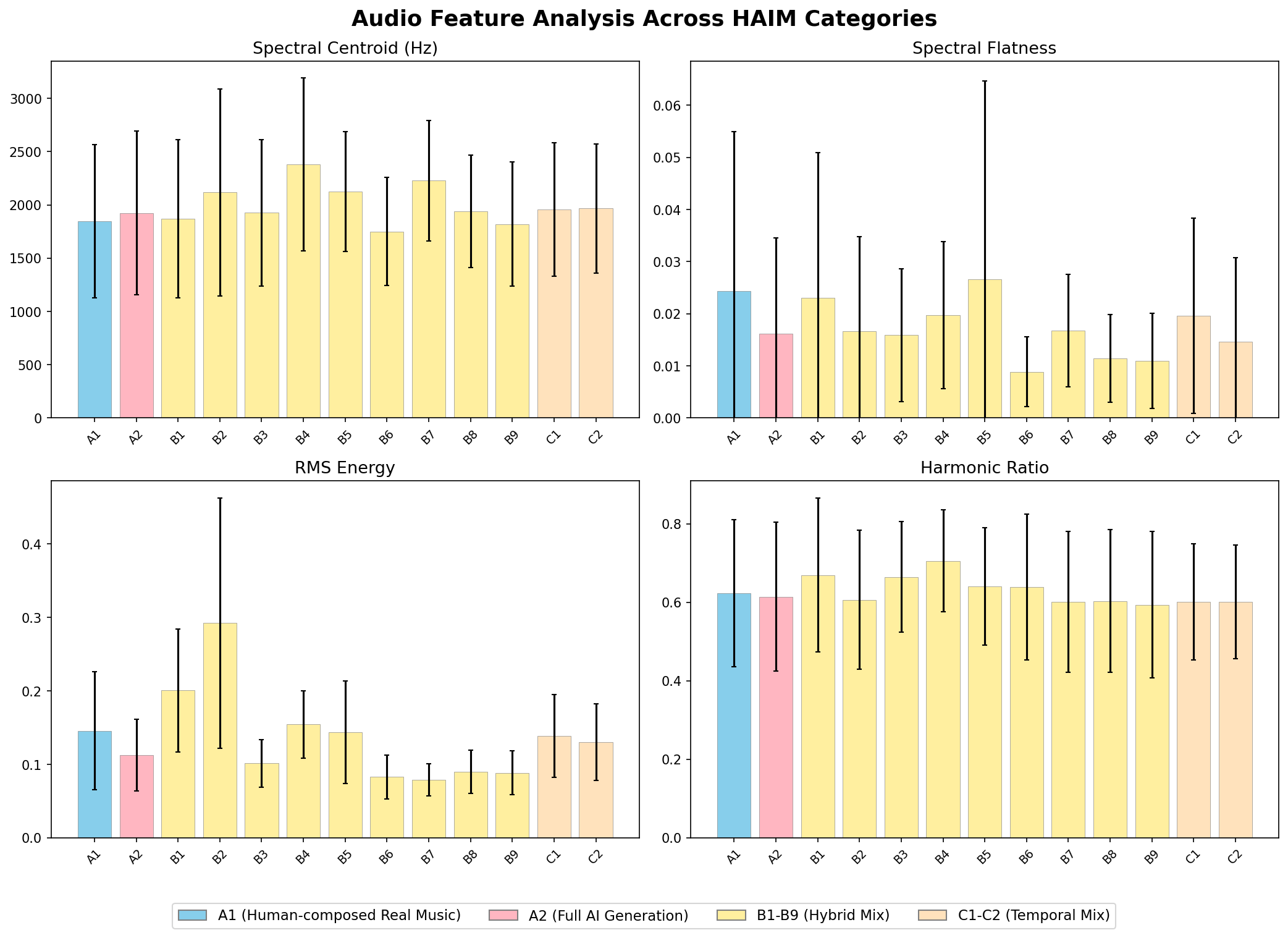}
  \caption{Spectral feature statistics across all HAIM categories. Bar heights indicate per-track means; error bars show standard deviations.}
  \label{fig:spectral}
  
  \vspace{12pt} 
  \small 
  \renewcommand{\arraystretch}{1.1} 
  \setlength{\tabcolsep}{5.5pt}
  \begin{tabular}{ccccccc} 
  \toprule
  \textbf{Category} & \textbf{Centroid (Hz)} & \textbf{Flatness} & \textbf{RMS} & \textbf{Harm. R.} & \textbf{Bandwidth (Hz)} & \textbf{Contrast (dB)} \\
  \midrule
  A1 & 1847 $\pm$ 720 & .024 $\pm$ .031 & .146 $\pm$ .080 & .624 $\pm$ .188 & 1982 $\pm$ 591 & 23.1 $\pm$ 2.1 \\
  A2 & 1924 $\pm$ 770 & .016 $\pm$ .018 & .113 $\pm$ .049 & .614 $\pm$ .190 & 2125 $\pm$ 562 & 23.2 $\pm$ 1.7 \\
  \midrule
  B1 & 1869 $\pm$ 743 & .023 $\pm$ .028 & .201 $\pm$ .084 & .670 $\pm$ .196 & 2001 $\pm$ 613 & 23.1 $\pm$ 2.2 \\
  B2 & 2117 $\pm$ 970 & .017 $\pm$ .018 & .292 $\pm$ .170 & .606 $\pm$ .177 & 2416 $\pm$ 626 & 23.0 $\pm$ 1.6 \\
  B3 & 1925 $\pm$ 686 & .016 $\pm$ .013 & .102 $\pm$ .033 & .665 $\pm$ .141 & 2228 $\pm$ 533 & 24.8 $\pm$ 1.6 \\
  B4 & 2379 $\pm$ 810 & .020 $\pm$ .014 & .154 $\pm$ .046 & .706 $\pm$ .130 & 2477 $\pm$ 550 & 24.7 $\pm$ 1.6 \\
  B5 & 2126 $\pm$ 561 & .027 $\pm$ .038 & .144 $\pm$ .070 & .641 $\pm$ .149 & 2286 $\pm$ 410 & 22.9 $\pm$ 1.7 \\
  B6 & 1750 $\pm$ 506 & .009 $\pm$ .007 & .083 $\pm$ .030 & .639 $\pm$ .185 & 2182 $\pm$ 411 & 24.0 $\pm$ 1.5 \\
  B7 & 2227 $\pm$ 566 & .017 $\pm$ .011 & .079 $\pm$ .022 & .601 $\pm$ .180 & 2493 $\pm$ 361 & 24.3 $\pm$ 1.4 \\
  B8 & 1939 $\pm$ 528 & .011 $\pm$ .008 & .090 $\pm$ .030 & .603 $\pm$ .182 & 2325 $\pm$ 387 & 24.5 $\pm$ 1.3 \\
  B9 & 1819 $\pm$ 582 & .011 $\pm$ .009 & .089 $\pm$ .030 & .594 $\pm$ .187 & 2218 $\pm$ 441 & 24.0 $\pm$ 1.5 \\
  \midrule
  C1 & 1957 $\pm$ 628 & .020 $\pm$ .019 & .139 $\pm$ .056 & .601 $\pm$ .148 & 2244 $\pm$ 535 & 23.8 $\pm$ 1.6 \\
  C2 & 1967 $\pm$ 607 & .015 $\pm$ .016 & .130 $\pm$ .052 & .601 $\pm$ .145 & 2270 $\pm$ 511 & 24.0 $\pm$ 1.5 \\
  \bottomrule
  \end{tabular}
  \captionof{table}{Per-category spectral feature statistics. Centroid, Bandwidth, and Contrast are reported with means and standard deviations.}
  \label{tab:spectral}
\end{figure*}

%--- B: Component Hybrids ---%
\paragraph{B5: AI Vocal Cover.}
AI vocal covers are sourced from YouTube, where creators use voice-cloning tools to replace vocal performances in human-produced songs. Metadata was collected via the YouTube Data API using bilingual keywords (e.g., ``AI cover,'' ``AI \begin{CJK}{UTF8}{mj}커버\end{CJK}'') with a post-2023 filter. Claude LLM agents performed title- and keyword-level integrity checks, and audio was constrained to between 30\,s and 7\,min\,30\,s.

\paragraph{B6: Human Lyrics with AI Generation.}
ACE-Step-1.5 is conditioned on human-written lyrics from the K-pop lyrics dataset~\citep{kim2024k}. We used 1,000 pairs of original Korean lyrics and their English translations, yielding 2,000 tracks where only the lyricist role is human.

%--- B: AI Editing from Human Sources ---%
\paragraph{B7: AI Variation.}
ACE-Step-1.5's variation mode is applied to 2,000 human tracks (SONICS), modifying acoustic texture---timbre, instrumentation density---while preserving the underlying harmonic and structural content.

\paragraph{B8: AI Edit.}
ACE-Step-1.5's edit mode takes a human song as an audio reference and generates a new track that preserves the style and arrangement of the original, simulating a workflow where producers feed a reference to an AI system.

\paragraph{B9: AI Repaint.}
ACE-Step-1.5's repaint mode generates entirely new compositions around reference masking, employing a distinct generation pipeline that progressively refines the output.

%--- C: Temporal MixSet ---%
\paragraph{C1: Concatenation.}
Human segments (MTG-Jamendo) and AI segments (ACE-Step-1.5) of 15--60\,s are concatenated with a brief silent gap ($[0.1,\,0.5]$\,s). Each track consists of multiple segments, with segment order randomized per track.

\paragraph{C2: Crossfade.}
Segments of 20--70\,s are blended with a triangular window of 1--5\,s, creating smooth music-mix transitions. Each track is composed of multiple segments.

\begin{figure*}[!t]
  \centering
  \includegraphics[width=\textwidth]{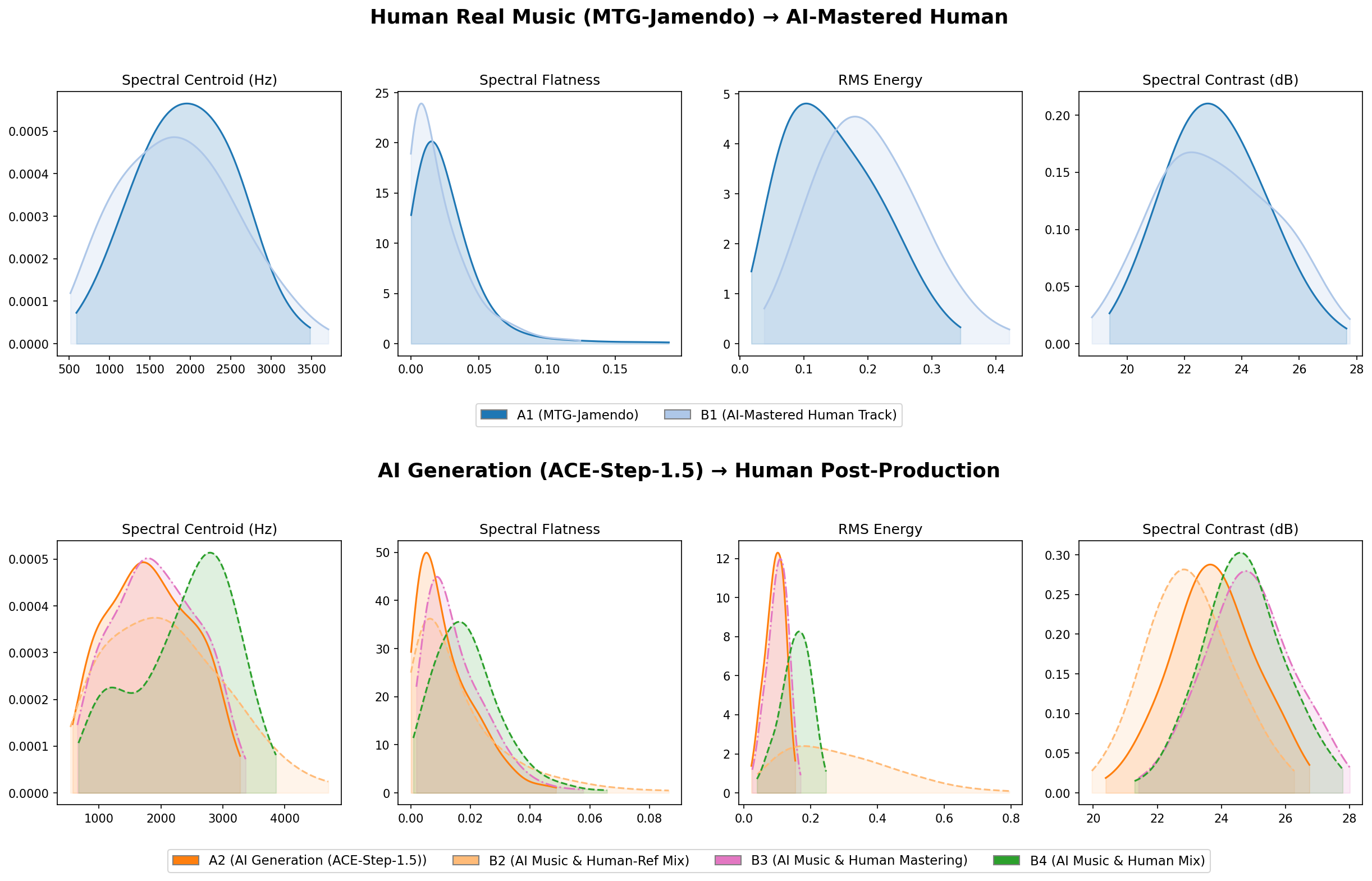}
  \caption{Kernel density estimates of seven acoustic features comparing two post-production pipelines. Left column: human music before and after AI mastering (A1\,$\to$\, B1). Right column: AI-generated music before and after human-referenced AI style transfer (A1\,$\to$\, B2), and human post-production (A2\,$\to$\, B3, B4).}
  \label{fig:kde_mastering}
\end{figure*}

\section{Dataset Analysis}
To characterize the acoustic landscape of HAIM prior to training, we analyze six low-level audio features. Spectral centroid and bandwidth represent the perceived brightness and frequency spread, respectively. Spectral flatness quantifies tonality versus noisiness, while spectral contrast measures peak-to-valley differences across sub-bands. RMS energy serves as a loudness proxy, and the harmonic ratio---derived via harmonic--percussive source separation---indicates the proportion of pitched versus transient content. Per-track statistics are reported in Figure~\ref{fig:spectral}.

Spectral centroid is broadly stable across categories, suggesting that AI generation preserves global frequency balance. Spectral flatness, however, is consistently lower for AI tracks (A2) than human recordings (A1), reflecting the smoother spectral envelopes of neural audio codecs~\citep{afchar2024detecting}. This gap narrows in hybrid categories with human post-production (B3, B4), indicating that professional mastering partially restores spectral irregularity.

RMS energy separates production pipelines rather than generation methods: AI-mastered (B1) and style-transferred (B2) tracks are substantially louder, while reference-conditioned categories (B6--B9) are quieter on average. Harmonic ratio remains broadly consistent, with a slight elevation in mastered tracks---likely a byproduct of noise reduction during mastering. Figure~\ref{fig:kde_mastering} provides a more detailed view of these distributional shifts, comparing the kernel density estimates of acoustic features before and after each post-production pipeline.

%===============================================================================
\section{Model}\label{sec:method}
%===============================================================================
Existing detectors produce a single binary label per track, which suffices for whole-track human versus AI classification but overlooks the granular role-level structure that defines HAIM's hybrid categories. To address this, we utilized Fusion Segment Transformer (FST)~\citep{kim2026fusion} model for the AI tracking task. FST is an architecture designed to fuse segment-level features via a transformer aggregator for track-level prediction. We modify the original FST by (1) replacing the audio encoder with MuQ~\citep{zhu2025muq}, a large-scale pre-trained music understanding backbone, and (2) removing the beat-tracking branch to reduce computational overhead. We refer to this modified variant as \textbf{MuQ-FST}.

\subsection{Architecture}\label{sec:architecture}
\paragraph{Stage 1: Segment Detection.}
An input track is divided into non-overlapping segments of 30\,s, each resampled to 24\,kHz and encoded by MuQ~\citep{zhu2025muq}, a 333M-parameter transformer pre-trained via self-supervised learning on large-scale music data. MuQ comprises 13 layers and produces hidden states. We fine-tune the last 6 layers while freezing earlier ones, yielding approximately 145M trainable parameters.

\paragraph{Stage 2: Full Song Detection.}
Segment-level representations are aggregated into a track-level prediction using the same fusion transformer as FST, with the beat-tracking branch removed. While this approach may provide less structural information, it still enables full-track analysis and reduces computational overhead.
% ---it added significant overhead.
% without proportional gains when paired with MuQ.

%===============================================================================
\section{Experiments}\label{sec:experiments}
%===============================================================================

\subsection{Settings}\label{sec:systems}

\begin{table*}[!t]
\centering
\small
\begin{tabular}{l|cc|ccccccccc|cc}
\toprule
& \multicolumn{2}{c|}{\textbf{Baselines}} & \multicolumn{9}{c|}{\textbf{Hybrid}} & \multicolumn{2}{c}
{\textbf{Temporal}} \\
\midrule
\textbf{Detector} & A1 & A2 & B1 & B2 & B3 & B4 & B5 & B6 & B7 & B8 & B9 & C1 & C2 \\
\rowcolor{gray!15}
\multicolumn{14}{l}{\textit{Open-source detectors}} \\
Deezer & 1.7 & 74.2 & 3.7 & 99.7 & 100.0 & 100.0 & 69.7 & 100.0 & 100.0 & 100.0 & 99.9 & 92.8 & 91.9 \\
SpecTTTra & 14.6 & 44.8 & 19.1 & 60.3 & 61.1 & 78.3 & 47.4 & 66.5 & 60.9 & 68.4 & 62.6 & 30.4 & 33.1 \\
CLAM & 14.7 & 65.2 & 12.8 & 83.7 & 92.5 & 79.9 & 59.2 & 97.1 & 88.2 & 97.9 & 95.5 & 63.0 & 64.4 \\
FST & 5.0 & 59.8 & 0.9 & 2.8 & 4.4 & 2.2 & 40.8 & 99.8 & 11.0 & 100.0 & 99.6 & 1.8 & 1.6 \\
\midrule
\rowcolor{cyan!15}
MuQ-FST (Ours) & \textbf{0.1} & \textbf{99.8} & 52.0 & 99.8 & 100.0 & 100.0 & 41.7 & 100.0 & 99.9 & 100.0 & 99.9 & 79.7 & 75.2 \\
\bottomrule
\end{tabular}
\caption{\textbf{AI Detection Rate (\%) across HAIM categories.} For A1, lower is better (human tracks misclassified as AI). For A2, higher is better (AI tracks correctly detected). For hybrid categories (B--C), the rate is descriptive---there is no single ``correct'' value. Detectors differ in training data and supervision (Section~\ref{sec:systems}); the A2 column reports the unweighted mean over the six generators in Table~\ref{tab:source_results}, whereas A1 is a track-level rate, so cross-column comparisons should be made with care.}
\label{tab:main_results}
\end{table*}
We evaluate Deezer's model with fourier transform~\citep{afchar2025fourier}, SpecTTTra~\citep{rahman2025sonics} ($\gamma$-120\,s configuration), CLAM~\citep{crosvila2025momclam}, Fusion Segment Transformer (FST)~\citep{kim2026fusion}, and our MuQ-FST. For training, we used AdamW (lr $5{\times}10^{-6}$, weight decay $0.01$, gradient clipping $0.5$), effective batch size 64, and early stopping (patience 50) on validation loss.

\paragraph{Binary detection.}
MuQ-FST is trained on A1 (Full Human) and A2 (Full AI) with cross-entropy loss, class-balanced sampling, and audio augmentations (Gaussian noise, time shifting, gain perturbation, MP3 compression at 128--320\, kbps, polarity inversion, clipping distortion). For Deezer and CLAM, only the feature extractor is publicly released; we trained a linear classifier on the same data. SpecTTTra and FST were evaluated using pretrained checkpoints trained on SONICS. As a result, the six A2 generators are in-distribution for MuQ-FST but largely out-of-distribution for the SONICS-pretrained checkpoints; cross-system comparisons should therefore be interpreted as indicative rather than a controlled head-to-head, and the per-generator breakdown in Table~\ref{tab:source_results} should be read with the in-distribution versus OOD status of each system in mind.

We report the AI Detection Rate: the percentage of tracks classified as AI-generated. For A1, lower is better; for A2, higher is better. For hybrid groups (B, C), the rate is purely descriptive---no single correct value exists.

\paragraph{Multilabel tracking.}
The binary head is replaced by a multilabel head trained with binary cross-entropy loss across four roles. Ground-truth labels follow Table~\ref{tab:role_taxonomy}. Group C, Category B2, and B7--B9 are excluded from training since role assignments vary within a single track.

For the multilabel task (MuQ-FST only), we report per-role AI probability and exact match ratio (fraction of tracks where all four role predictions are correct).

\subsection{Binary Detection Results}\label{sec:binary_results}
Table~\ref{tab:main_results} reports the AI Detection Rate across all HAIM categories, and Table~\ref{tab:source_results} further disaggregates A2 performance by generation source. We emphasize that only A1 and A2 have definitive ground truth (human and AI, respectively); for hybrid (B) and temporal mix (C) categories, the reported rates are descriptive and reflect each detector's sensitivity to AI-involved content rather than correctness.

\begin{table}[h]
\centering
\small
\resizebox{\columnwidth}{!}{%
\begin{tabular}{l|cccccc}
\toprule
\textbf{Detector} & \textbf{Suno} & \textbf{ACE} & \textbf{MusicGen} & \textbf{Udio} & \textbf{Mureka} & \textbf{Lyria} \\
\midrule
Deezer    & 97.4 & \textbf{100.0} & 92.4 & 92.0 & 54.9 & 8.3 \\
CLAM & 44.6 & 96.3 & \textbf{100.0} & 43.3 & 66.1 & 41.0 \\ 
FST & 96.4 & 85.6 & 73.3 & 86.5 & 2.5 & 9.6 \\
SpecTTTra & 51.3 & 57.8  & 51.9 & 51.1 & 38.9 & 18.0 \\

\midrule
\rowcolor{cyan!15}
MuQ-FST (Ours) & \textbf{99.9} & 99.8 & \textbf{100.0} & \textbf{99.8} & \textbf{99.6} & \textbf{100.0} \\
\bottomrule
\end{tabular}%
}
\caption{AI Detection Rate (\%) for A2 category across various generators. For MuQ-FST all six generators are in-distribution, whereas for the SONICS-pretrained SpecTTTra and FST, ACE-Step, MusicGen, Mureka, and Lyria are out-of-distribution; the gap should be read in that light.}
\label{tab:source_results}
\end{table}

\begin{table*}[t]
\centering
\small
\begin{tabular}{l|cccc|c}
\toprule
\textbf{Category} & \textbf{Composer} & \textbf{Lyricist} & \textbf{Vocalist} & \textbf{Engineer} & \textbf{Exact Match} \\
\midrule
A1 --- Full Human           & 0.0 & 0.0 & 0.0 & 0.0 & 100.0\% \\
A2 --- Full AI               & \textbf{99.9} & \textbf{99.9} & \textbf{99.9} & \textbf{99.9} & 99.9\% \\
\midrule
B1 --- AI Master-Human       & 0.0 & 0.0 & 0.0 & \textbf{100.0} & 100.0\% \\
B2 --- AI \& Human-Ref Mix    & \textbf{89.4} & \textbf{89.3} & \textbf{94.3} & \textbf{84.9}$^\dagger$ & 80.1\% \\
B3 --- AI \& Human Mastering       & \textbf{100.0} & \textbf{100.0} & \textbf{100.0} & 0.0 & 100.0\% \\
B4 --- AI \& Human Mix           & \textbf{100.0} & \textbf{100.0} & \textbf{100.0} & 0.0 & 100.0\% \\
B5 --- AI Vocal Cover        & 1.1 & 1.1 & \textbf{98.9} & \textbf{99.0} & 98.1\% \\
B6 --- Human Lyrics w/ AI Gen     & \textbf{100.0} & 100.0 & \textbf{100.0} & \textbf{99.9} & 0.0\% \\
\midrule
B7 --- AI Variation          & \textbf{99.3}$^\dagger$ & \textbf{99.3}$^\dagger$ & \textbf{99.7}$^\dagger$ & \textbf{83.4}$^\dagger$ & 83.6\% \\
B8 --- AI Edit               & \textbf{99.2}$^\dagger$ & \textbf{99.1}$^\dagger$ & \textbf{99.9}$^\dagger$ & \textbf{85.0}$^\dagger$ & 85.0\% \\
B9 --- AI Repaint            & \textbf{99.1}$^\dagger$ & \textbf{99.1}$^\dagger$ & \textbf{99.6}$^\dagger$ & \textbf{83.9}$^\dagger$ & 84.3\% \\
\midrule
C1 --- Concatenation         & 79.1$^\bullet$ & 79.1$^\bullet$ & 85.9$^\bullet$ & 76.9$^\bullet$ & --- \\
C2 --- Crossfade             & 63.7$^\bullet$ & 63.7$^\bullet$ & 81.0$^\bullet$ & 73.8$^\bullet$ & --- \\
\bottomrule
\end{tabular}
\caption{\textbf{Per-category multilabel predictions by our MuQ-FST (validation set).} Each cell shows the mean predicted AI probability (\%) for each production role. \textbf{Bold} indicates ground-truth AI roles; \textbf{AI$^\dagger$} denotes AI generation conditioned on human source material.}
\label{tab:multilabel_category}
\end{table*}
Among open-source systems, Deezer achieves the lowest false positive rate on real music (A1: 1.7\%) and strong detection on full AI content (A2: 74.2\%), but exhibits notable source dependence---Lyria Pro-3 evades almost entirely (8.3\%). SpecTTTra-$\gamma$ shows weaker separation overall, with no generation source exceeding 58\%. Notably, Lyria Pro~3 proves consistently difficult to detect across all evaluated systems---Deezer (8.3\%), SpecTTTra (18.0\%), FST (9.6\%), and CLAM (41.0\%)---which may reflect either qualitatively different generation artifacts or, for the SONICS-pretrained systems, its out-of-distribution status. MuQ-FST achieves near-perfect detection on full AI content (A2: 99.6\%) with a false positive rate (A1: 0.1\%), though we note that all six generators are in-distribution for MuQ-FST.

The hybrid categories reveal how each detector responds to varying degrees of AI involvement. AI mastering of human tracks (B1) triggers minimal responses across most systems, while human post-processing of AI tracks (B2--B4) preserves high AI rates for Deezer, indicating that neural generation fingerprints persist through conventional mastering chains. Reference-conditioned categories (B7--B9) and temporal mix set tracks (C1--C2) further illustrate divergent detector behaviors depending on the nature and proportion of AI content.

\subsection{Multilabel AI Music Tracking}\label{sec:multilabel}
This section evaluates MuQ-FST on the multilabel tracking task: given a track, predict which of the four production roles involves AI. The detailed results for this task are summarized in Table~\ref{tab:multilabel_category}.

% This is the core contribution that operationalizes the tracking formulation defined in Section~\ref{sec:intro}. 

% The multilabel results demonstrate that MuQ-FST performs remarkably well on the \emph{detection} task: across all categories, the model reliably distinguishes AI-involved roles from purely human ones, achieving near-perfect exact match rates in most cases.

B3 and B4 yield results closely aligned with their ground truth, correctly predicting AI involvement in Composer, Lyricist, and Vocalist while assigning near-zero probability to Engineer. However, since all B3 and B4 tracks were processed through the same mastering and mixing template, there is a risk that the model has overfit to template-specific artifacts rather than learning generalizable cues of human post-production. Incorporating a wider variety of human mixing and mastering styles in future work would help disentangle this confound. Nevertheless, the model's predictions on B3 and B4 are clearly distinct from those on A1, which contains the pre-mixed versions of the same underlying Ace-Step tracks, suggesting that the model is capturing meaningful acoustic differences introduced by the human mixing and mastering stage.
\begin{figure*}[!t]
\centering
\includegraphics[width=\textwidth]{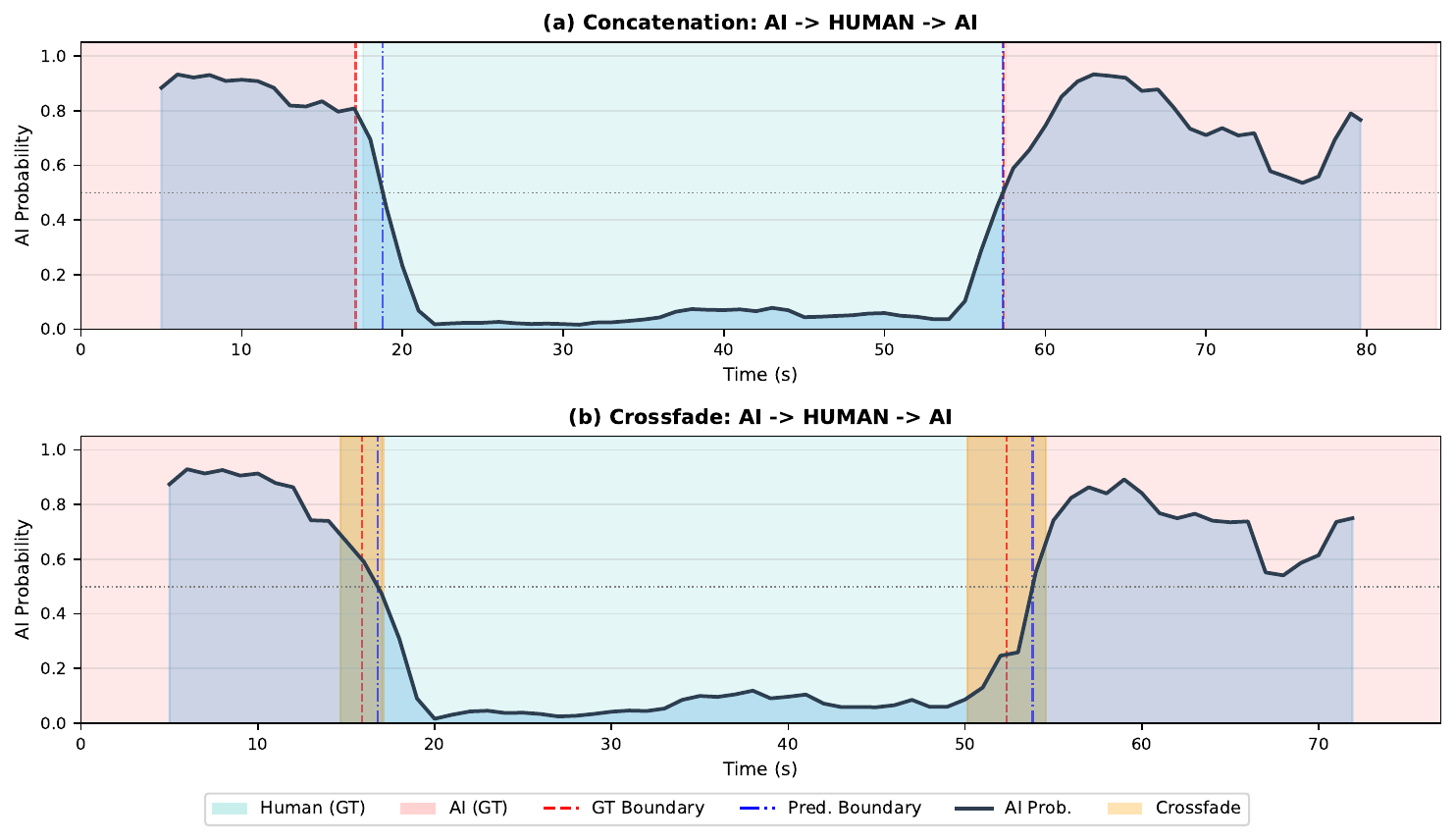}
\caption{Temporal AI probability curves (10\,s window, 1\,s hop) for representative C1 (top) and C2 (bottom) tracks. Green/red shading: ground-truth human/AI segments. Dashed red: ground-truth boundaries; dash-dotted blue: predicted boundaries. Orange: crossfade zone (C2 only).}
\label{fig:temporal_heatmap}
\end{figure*}

The predicted probabilities for Composer and Lyricist are nearly identical within each category, indicating that the model cannot yet disentangle these roles from the audio signal alone, especially in Lyricist. As shown in B6, where lyrics are the only human-contributed role, the model still predicts a high AI probability for the Lyricist role, even under explicit supervised training. This suggests that purely acoustic representations are insufficient for capturing linguistic authorship. Consequently, integrating auxiliary modules, such as automatic speech recognition (ASR) or lyric-aware text encoders~\citep{frohmann2025ai}, may be necessary to reliably attribute the Lyricist role.

%===============================================================================
\subsection{Temporal AI Localization}\label{sec:temporal}
%===============================================================================
Since MuQ-FST performs segment-level analysis in Stage~1, the model naturally produces per-segment AI probabilities that can be repurposed for temporal localization without any additional training. We exploit this property to investigate whether AI-generated regions can be localized within temporally hybrid tracks, using C1 (concatenation) and C2 (crossfade) from the HAIM benchmark.

A sliding window of size $w{=}10$\,s with hop $h{=}1$\,s is applied to each track, producing an AI probability curve $p(t)$. Each window is zero-padded to 30\,s to match the expected input length of the MuQ encoder. Boundaries are detected at threshold crossings $p(t) = \tau$, with linear interpolation between adjacent windows for sub-hop precision.

We report boundary \textbf{Precision} (P), \textbf{Recall} (R), and \textbf{F1} ($\uparrow$), where a predicted boundary is a true positive if it falls within tolerance $\delta$ seconds of a ground-truth boundary. \textbf{MBE} (Mean Boundary Error, $\downarrow$) is the mean absolute displacement between matched pairs:
\begin{equation}
  \text{MBE} = \frac{1}{|\text{TP}|}\sum_{(b_i,\hat{b}_i)\in\text{TP}} |b_i - \hat{b}_i|.
\end{equation}
\textbf{Segment IoU} ($\uparrow$) measures the frame-level overlap between predicted and ground-truth AI regions. The probability curve $p(t)$ is thresholded at $\tau$ to produce a binary mask $\mathbf{m}_{\text{pred}} \in \{0,1\}^{N}$, and the ground-truth AI regions yield $\mathbf{m}_{\text{gt}} \in \{0,1\}^{N}$,
\begin{equation}
  \text{IoU} = \frac{\sum_{n=1}^{N} \mathbf{m}_{\text{pred}}[n] \;\land\; \mathbf{m}_{\text{gt}}[n]}{\sum_{n=1}^{N} \mathbf{m}_{\text{pred}}[n] \;\lor\; \mathbf{m}_{\text{gt}}[n]}.
\end{equation}

\paragraph{Results.}

Figure~\ref{fig:temporal_heatmap} shows the AI probability curve
with the 10\,s window on representative tracks.
The predicted boundaries closely align with ground-truth annotations,
and the AI probability drops sharply to near zero in human segments.

\begin{table}[h]
\centering
\small
\begin{tabular}{ll|ccc|cc}
\toprule
& \textbf{$w$ / $h$} & \textbf{P $\uparrow$} & \textbf{R $\uparrow$} & \textbf{F1 $\uparrow$}
  & \textbf{MBE $\downarrow$} & \textbf{IoU $\uparrow$} \\
\midrule
\multirow{4}{*}{\small\textbf{C1}}
& 30\,s / 3\,s & .364 & .284 & .305 & 2.2 & .554 \\
& 20\,s / 2\,s & .570 & .537 & .542 & 2.4 & .667 \\
& 10\,s / 1\,s & \textbf{.897} & .948 & \textbf{.914} & 2.2 & .764 \\
& 5\,s / 1\,s  & .672 & \textbf{.975} & .760 & \textbf{1.4} & \textbf{.778} \\
\midrule
\multirow{3}{*}{\small\textbf{C2}}
& 30\,s / 1\,s & .370 & .282 & .299 & 2.6 & .528 \\
& 10\,s / 1\,s & \textbf{.759} & .839 & \textbf{.783} & 2.4 & \textbf{.672} \\
& 5\,s / 1\,s  & .679 & \textbf{.962} & .765 & \textbf{2.0} & .668 \\
\bottomrule
\end{tabular}
\caption{Temporal localization with varying window sizes
($n{=}200$ tracks, $\tau{=}0.5$, $\delta{=}5$\,s).
All windows are zero-padded to 30\,s for the MuQ encoder.
$\uparrow$: higher is better; $\downarrow$: lower is better.}
\label{tab:window_ablation}
\end{table}

Table~\ref{tab:window_ablation} summarizes the localization performance across different window sizes. With the 10\,s window, boundary F1-score reaches 0.914 on C1 and 0.783 on C2, a threefold improvement over the 30\,s baseline. These results demonstrate that temporal AI localization emerges as a zero-shot capability of the segment-level MuQ encoder.
Future work could combine predictions from multiple window sizes
through multi-scale fusion, or employ dedicated temporal architectures
trained with boundary supervision.
%===============================================================================
\section{Discussion}\label{sec:discussion}
%===============================================================================

\subsection{Limitations}

Our multilabel formulation does not fully capture the diverse and nuanced ways human and AI contributions are combined in real-world music production. For example, the current formulation operates at the Composer level, merging the roles of Tracker (arrangement/instrumentation) and Top-liner (melody writing) into a single label. In practice, these roles can be independently performed by humans or AI—a human may provide a hummed melody while an AI completes the full arrangement, or conversely, a human may produce the instrumental while an AI generates the vocal melody on top. A more flexible, scenario-driven modeling approach beyond rigid label-based representations could better address these complexities.

A further limitation concerns generator coverage and the resulting confounds. Most of Group B is generated by a single backbone (ACE-Step-1.5), so a detector trained on these categories may learn generator-specific fingerprints rather than role-level cues; the in-distribution status of the A2 generators for MuQ-FST similarly limits how far the binary-detection comparison in Tables~\ref{tab:main_results} and~\ref{tab:source_results} can be read as a controlled head-to-head. Replicating the hybrid categories with additional generators and adopting held-out-generator evaluation would help disentangle role-level signal from generator identity.

The class distribution across categories is also highly imbalanced: the two baseline groups are drawn from large-scale sources (A1: 94K, A2: 55K), whereas the hybrid and temporal-mix categories rely on manual processing or single-generator pipelines and are far smaller (typically 2K--6K). This imbalance reflects the cost of constructing the hybrid categories rather than a design choice, and scaling them is left to future work.
 
 The hybrid categories B3 and B4 were constructed using a single mastering and mixing template, which may cause the model to overfit to template-specific artifacts rather than learning generalizable cues of human post-production. Because a fixed plugin chain and stereo configuration were applied uniformly across tracks, these categories should be read as a controlled, reproducible post-production condition rather than as representative of the full diversity of human mixing and mastering practice. Incorporating a wider diversity of mixing engineers, DAW environments, and mastering chains would better reflect real-world variability. The MixSet (C1, C2) covers only simple concatenation and crossfading; more sophisticated temporal manipulations such as AI phrase replacement within a vocal take or AI-assisted song restructuring are not addressed. 

Additionally, the FST-MuQ-based experiments were conducted under a relatively uniform detection setup. Only the label definitions were adapted to approximate different tracking tasks compared to existing detection tasks. There are diverse views in the HAIM dataset benchmark, such as agent-level attribution or temporal boundary detection, but the training objectives and architectural design remained largely unchanged. 

% Additionally, the FST-MuQ-based experiments were conducted under a relatively uniform detection setup. Only the label definitions were adapted to approximate different tracking tasks compared to existing detection tasks. There are many various view in HAIM dataset benchmark like agent-level attribution or temporal boundary detection, but the training objectives and architectural design remained largely unchanged. 
\subsection{Future Work}

Expanding the HAIM dataset with additional generation sources, newer model versions, and more diverse human mixing and mastering workflows is a natural next step. The current thirteen categories cover representative production scenarios, but real-world AI integration is far more varied---AI-assisted arrangement, AI chord suggestion, AI-driven sound design, and partial stem regeneration are all emerging practices that warrant dedicated categories. Scaling the number of tracks per category would strengthen the benchmark's statistical reliability.

For the Lyricist role, integrating auxiliary modalities such as automatic speech recognition (ASR) or lyric-aware text encoders is a promising direction, as our results suggest that purely acoustic representations are insufficient for capturing linguistic authorship. Incorporating stem-level detection approaches—such as those based on source separation—could further enhance the framework. Moreover, the integration of music understanding models, potentially combined with LLM-based reasoning, also presents a promising direction for capturing more complex and fine-grained patterns in music production.

HAIM is also closely related to adversarial robustness research. Some individuals use post-processing techniques to make AI-generated music appear human-made. The hybrid categories—particularly human mastering of AI tracks (B3, B4) and reference-conditioned generation (B7–B9)—already simulate realistic post-processing conditions. As these categories can unintentionally degrade detector performance, they serve as a valuable testbed for evaluating robustness. Future work may extend this benchmark by introducing intentional adversarial attacks, enabling a systematic study of the evolving detection–evasion arms race.

%===============================================================================
\section{Conclusion}\label{sec:conclusion}
%===============================================================================

The landscape of AI-assisted music production is rapidly diversifying, yet current detection systems reduce this complexity to a binary label. We propose \emph{AI Tracking} as a broader framework that moves beyond binary classification to characterize the nature and extent of AI involvement across production roles.

As a first step, we introduce HAIM, a dataset with structured role-level annotations. Using MuQ-FST, we demonstrate that multilabel agent classification, predicting AI involvement across Composer, Lyricist, Vocalist, and Engineer roles. We further explore temporal analysis to capture when AI contributions occur within a track, providing an additional layer of interpretability beyond static labeling.

Building on these results, AI Tracking should operate at multiple granularities. This includes explainable AI (XAI) methods for interpretable evidence, track-level detection identifying which stems are AI-generated, music pattern-level analysis for specific motifs or chord progressions, and MIDI-level detection analyzing symbolic representations. These different levels provide complementary views of AI involvement, from identifying whether a track contains AI-generated content to pinpointing how and where it contributes.

\section{Ethics}
All human music is sourced from public-domain, Creative Commons, or licensed collections. AI content was generated via paid subscriptions or open-source models under their respective licenses. AI vocal covers (B5) are sourced from publicly available YouTube uploads. No human subjects were involved. AI-generated tracks are research data, not intended for distribution as music. Among the released data, only ACE-Step-based generated tracks and their derived categories (B3, B4, B7--B9) are available under the MIT License. All remaining data are strongly recommended for research purposes only.

%%%%%%%%%%%%%%%%%%%%%%%%%%%%%%%%%%%%%%%%%%%%%%%%%%%%%%%%%%%%%%%%%%%%%%%%%%%%%%%%
% Bibliography
%%%%%%%%%%%%%%%%%%%%%%%%%%%%%%%%%%%%%%%%%%%%%%%%%%%%%%%%%%%%%%%%%%%%%%%%%%%%%%%%

\bibliography{TISMIRtemplate}

\end{document}